\documentclass[12pt]{amsart}
\usepackage{amsmath,amssymb,txfonts}
\usepackage{amssymb}
\usepackage{amsmath}
\usepackage{mathrsfs}
\usepackage{amsmath,amssymb}

\newtheorem{theorem}{Theorem}[section]
\newtheorem{lemma}[theorem]{Lemma}

\newtheorem{corollary}[theorem]{Corollary}
\theoremstyle{definition}
\newtheorem{definition}[theorem]{Definition}

\theoremstyle{remark}
\newtheorem{remark}[theorem]{Remark}

\def\R{\mbox{\rm{I}}\!\mbox{\rm{R}}}

\def\<{\leq}             \def\>{\geq}

\numberwithin{equation}{section}

\usepackage{color}

\begin{document}

\title[Ill-posedness of Navier-Stokes equations]
{Ill-posedness of Navier-Stokes equations and critical Besov-Morrey spaces}


\author[Qixiang Yang]{Qixiang Yang}
\address{
School of Mathematics and Statistics, Wuhan University, Wuhan, 430072, China.
}
\email{qxyang@whu.edu.cn}

\author[Haibo Yang]{Haibo Yang}
\address{Faculty of Mathematics and Statistics, Hubei University,
Wuhan, 430062, China.}
\email{yanghb97@qq.com}

\author[Huoxiong Wu]{Huoxiong Wu}
\address{School of Mathematical sciences, Xiamen University, Xiamen Fujian, 361005, China.
}
\email{huoxwu@xmu.edu.cn}


\thanks{This work was supported by the National Natural Science Foundation of China (No. 11571261, 11771358, 11871101).}


\date{}

\dedicatory{}


\begin{abstract}
The blow up phenomenon in the first step of the classical Picard's scheme is proved in this paper. For certain initial spaces $X\subseteq (\dot{B}^{-1,\infty}_{\infty})^{n}$, Bourgain-Pavlovi\'c and Yoneda proved the ill-posedness of the Navier-Stokes equations by showing the norm inflation in the spaces $L^{\infty}_{t}(X)$.
But Chemin and Gallagher said the space $(\dot{B}^{-1,\infty}_{\infty})^{n}$ seems to be optimal,
if one replace $L^{\infty}_{t}(X)$ by  some solution spaces best chosen.
In this paper, we use Meyer wavelets to construct different initial data in more general initial spaces
than those studied by Bourgain-Pavlovi\'c and Yoneda
and establish new ill-posedness result, which is independent of the choice of solution space.
Our result is a nice complement of the previous ill-posedness results on Navier-Stokes equations.
\end{abstract}

\maketitle


 \vspace{0.1in}

\section{Introduction and main result} 
\label{intro}
We consider the incompressible Navier-Stokes equations
\begin{equation}\label{eqn:ns}
\left\{\begin{array}{ll} \frac{\partial u} {\partial t}
-\Delta u + u \cdot \nabla u -\nabla p=0,
& \mbox{ in }  [0,T)\times \mathbb{R}^{n}; \\
\nabla \cdot u=0,
& \mbox{ in } [0,T)\times \mathbb{R}^{n}; \\
u|_{t=0}= u_0, & \mbox{ in } \mathbb{R}^{n};
\end{array}
\right.
\end{equation}
where $u(t,x)$ and $p(t,x)$ denote the velocity vector field and the pressure of fluid at the point $(t,x)\in [0,T)\times \mathbb{R}^{n}$ respectively. While $u_0$ is a given initial velocity vector field. Cannone \cite{C1} established the well-posedness of (\ref{eqn:ns}) for Besov spaces, Koch-Tataru \cite{KT} obtained the well-posedness  of (\ref{eqn:ns}) for ${\rm BMO}^{-1}$. Besov spaces and ${\rm BMO}^{-1}$ are special critical Besov-Morrey spaces. Later on, Li-Xiao-Yang \cite{LXY} showed the well-posedness for more general critical Besov-Morrey spaces.
In this paper, we consider the rest critical Besov-Morrey spaces, and we will show blow up phenomenon  of (\ref{eqn:ns}) in the first step of the classical Picard's scheme for these initial spaces.

The solutions of the above Cauchy problem can be obtained via the integral equation:
\begin{equation}\label{eqn:mildsolution}
u(t,x)= e^{t\Delta } u_0(x) - B(u,u)(t,x),
\end{equation}
where
\begin{equation}\label{eqn:HW}
\begin{cases}
B(u,u)(t,x)\equiv\int^{t}_{0} e^{(t-s) \Delta}
\mathbb{P}\nabla (u\otimes u) ds,\\
\mathbb{P}\nabla (u\otimes u)\equiv \sum\limits_{l}
\frac{\partial}{\partial x_{l}} (u_{l}u) -
\sum\limits_{l} \sum\limits_{l'} (-\Delta)^{-1} \frac{\partial}{\partial x_{l}} \frac{\partial}{\partial x_{l'}} \nabla (u_{l} u_{l'}).
\end{cases}
\end{equation}
The equation (\ref{eqn:mildsolution}) can be solved by a fixed-point method whenever the convergence
is suitably defined in certain function spaces. For any $l,\,l',\,l''\in \{1,\cdots,n\}$, denote
\begin{equation}\label{eqn:B}
\begin{cases}
B_l(u,v)(t,x)\equiv\int^{t}_{0} e^{(t-s)\Delta} \frac{\partial}{\partial x_l}
(u(s,x) v(s,x)) ds,\\
B_{l.l',l''}(u,u)(t,x)\equiv\int^{t}_{0} e^{(t-s)\Delta} (-\Delta)^{-1}\frac{\partial}{\partial x_l} \frac{\partial}{\partial x_{l'}} \frac{\partial}{\partial x_{l''}}
(u(s,x) v(s,x)) ds.
\end{cases}
\end{equation}
For $u_0$ belongs to some initial space $X^n= (X(\mathbb{R}^{n}))^{n}$, denote
\begin{equation}\label{eqn:it}
\begin{cases}
u^{(0)}(t,x) = e^{t\Delta} u_0,\\
u^{(j+1)} (t,x) = u^{(0)}(t,x) - B(u^{(j)}, u^{(j)})(t,x), \forall j=0,1,2,\cdots
\end{cases}
\end{equation}
where $e^{t\Delta}u_0$ belongs to some space $Y^{n}= (Y([0,T)\times \mathbb{R}^n))^{n}$.
To prove $$B(u,v) {\mbox {\rm \, are bounded from }} Y^n\times Y^n {\mbox {\rm \, to }} Y^n,$$
the traditional method is to prove the following conclusion is true:
\begin{equation}\label{eq:bb}\begin{array}{c}
B_{l}(u,v), B_{l,l',l''}(u,v) {\mbox {\rm \, are bounded from }} Y\times Y {\mbox {\rm \, to }} Y,\\
\forall l,l',l''\in \{1,\cdots,n\}.
\end{array}\end{equation}
The above iteration process convergence for $\|u_0\|_{X^n}$ small enough.
Such solutions of (\ref{eqn:mildsolution}) are called mild solutions of
(\ref{eqn:ns}). The notion of such a mild solution was pioneered by
Kato-Fujita \cite{KF} in 1960s. During the latest decades, many
important results about mild solutions to (\ref{eqn:ns}) have been
established. See, for example, Cannone \cite{C1, C2},
Germin-Pavlovic-Staffilani \cite{GPS}, Giga-Miyakawa \cite{GM}, Lemari\'e \cite{Lem, Lem1},
Kato \cite{Kat},  Koch-Tataru \cite{KT}, Wu \cite{W1,W2,W3,W4} and the references therein.

Morrey spaces were introduced in 1938 by Morrey \cite{Mo}. Many authors extended them to two kinds of oscillation spaces: Besov-Morrey spaces and Triebel-Lizorkin-Morrey spaces. They considered the well-posedness results for initial value in generalized Morrey spaces, see \cite{LXY, LiY, LY,YL} {\it etc}.
In this paper, we consider the ill-posedness in critical Besov-Morrey spaces. Choose
\begin{equation}\label{eq:phi}\begin{array}{c}
\phi(x)\in C^{\infty}_{0}(B(0,2n)) \mbox{ such that }\\ \phi(x) \mbox{ takes value } 1 \mbox{ on the ball } B(0,2\!\sqrt{n}).
\end{array}
\end{equation}
Let $Q(x_0,r)$ be a cube with sides parallel to the coordinate axes, centered at $x_0$ and with side length $r$, $\phi_{Q}(x)= \phi(\frac{x-x_0}{r})$, and $\mathfrak{C}= \{Q(x_0,r), x_0\in \mathbb{R}^{n}, r>0\}$. For $1\leq p,q\leq \infty$ and $\gamma_1,\gamma_2\in \mathbb{R}$, let $m_0= m^{\gamma_1,\gamma_2}_{p,q}$ be a positive constant large enough. For arbitrary function $f(x)$, let $S^{\gamma_1,\gamma_2}_{p,q,f}$ be the set of polynomial functions $P_{Q,f}(x)$  of order less than $m_0$.

Now we introduce the definition of Besov-Morrey spaces which includs Besov spaces,  Morrey spaces, ${\rm BMO}^{-1}$ and $Q$ spaces {\it etc} (see \cite{LY,YSY}).
\begin{definition}
Given $1\leq p,q\leq \infty$ and $\gamma_1\in \mathbb{R}, 0\leq \gamma_2\leq \frac{n}{p}$.
Besov-Morrey spaces $\dot{B}^{\gamma_1,\gamma_2}_{p,q}(\mathbb{R}^{n})$ are defined as follows:
$$\sup\limits_{Q\in \mathfrak{C}} |Q|^{\frac{\gamma_2}{n}-\frac{1}{p}} \inf_{P_{Q,f}\in S^{\gamma_1,\gamma_2}_{p,q,f}}
\|\phi_{Q}(f-P_{Q,f})\|_{\dot{B}^{\gamma_1,q}_{p}}<\infty.$$
\end{definition}

Critical spaces occupied a significant place for Navier-Stokes equations (\ref{eqn:ns}).
For the above oscillation spaces, according to Lemma \ref{le:2.5}, if $\gamma_1-\gamma_2=-1$, then they are critical spaces.
Further, one pays attention to the largest critical spaces.
According to Lemma \ref{lem:2.9},  $\dot{B}^{-1,0}_{p,q}  (1\leq p,q\leq \infty)$ are the relative large spaces among the above critical oscillation spaces.

For initial values in some critical initial spaces $X^{n}\subseteq (\dot{B}^{-1,\infty}_{\infty})^{n}$, one has proved norm inflation in solution spaces $Y^{n}=L^{\infty}([0,T), X^{n})$.  In fact, Bourgain-Pavlovi\'c \cite{BP} considered integral Bloch space $X= \dot{B}^{-1,\infty}_{\infty}$. For $q>2$, Yoneda \cite{Yo} considered Besov spaces $X= \dot{B}^{-1,q}_{\infty} $ and Triebel-Lizorkin spaces $X=\dot{F}^{-1,q}_{\infty}$ which are special Besov-Morrey spaces $X=\dot{B}^{-1,0}_{q,q}$.
Chemin and Gallagher \cite{CG} said the space $(\dot{B}^{-1,\infty}_{\infty})^{n}$ seems to be  optimal, if one replace $L^{\infty}([0,T), X^{n})$ by certain solution space well chosen.
For all the critical Besov-Morrey spaces, according to Lemmas \ref{lem:2.9} and \ref{le:2.7},
only $\dot{B}^{-1,q}_{\infty}(q>2)$ or $\dot{B}^{-1,0}_{2,q'}(q'>2)$ or $\dot{B}^{-1,0}_{p,p'}(p>2,p'\geq 1)$
can not be contained in ${\rm BMO}^{-1}$.
For these three classes of spaces, we will prove the equation (\ref{eq:bb}) is not true.
\begin{definition}
For initial data $u_0\in X^{n}$, if one can not find $Y^{n}$ such that $e^{t\Delta}u_0\in Y^{n}$ and the equation (\ref{eq:bb}) is  true, then we say
the classical Picard's process  can not be applied to the initial value space $X^{n}$ for the Navier-Stokes equations (\ref{eqn:ns}).
\end{definition}
We will show that the classical Picard's process can not be applied to the above three classes function spaces in this paper.
Let \begin{equation}\label{eq:e} \vec{e}= (1,\cdots, 1),\end{equation}
\begin{equation}\label{eq:g} g(t,x) = {\rm \, exp \,} \{- \frac{x^{2}} {4t}\}, \forall t>0, x\in \mathbb{R}^{n}.\end{equation}
Precisely, our main results can be formulated as follows..

\begin{theorem} \label{mthmain}
For any $\delta>0$, there exists $u_0=(u_1,u_2,0,\cdots, 0)^{t}$ satisfying

$(i)$\, ${\rm div\, } u_0=0$.

$(ii)$\, $\forall p>2, p'\geq 1, q>2,q'>2,$
$$\max\{\|u_0\|_{(\dot{B}^{-1,q}_{\infty})^{n}}, \|u_0\|_{(\dot{B}^{-1,0}_{2,q'})^{n}}, \|u_0\|_{(\dot{B}^{-1,0}_{p,p'})^{n}} \}\leq \delta.$$

$(iii)$\, $u_0$ belongs to $({\mathcal{S}}(\mathbb{R}^n))^{n}$ outside some ball: $$(1-\phi(x-\frac{3}{2}\vec{e} )) u_0\in ({\mathcal{S}}(\mathbb{R}^n))^{n}.$$

$(iv)$\, $\forall 0<t<1$,
\begin{equation}\label{eq:blowup}B_1(e^{t\Delta} u_1, e^{t\Delta } u_1) \notin {\mathcal{S}}'(\mathbb{R}^{n}).\end{equation}
In fact, for $0<t<1$, we have  $g(t,x)\in {\mathcal{S}}(\mathbb {R}^{n})$ and
\begin{equation}\label{eq:main}
\langle B_1 (e^{t\Delta} u_1, e^{t\Delta}u_1), g(t,x)\rangle = \infty.
\end{equation}
\end{theorem}

We remark that all the published ill-posedness results of (\ref{eqn:ns}) depended on the choice of solution spaces.
But our ill-posedness result does not depend on the choice of solution spaces.

\begin{remark}
(i) The meaning of ill-posedness in \cite{BP} and \cite{Yo} is different to which in this paper.
For any $\delta>0, T>0$, Bourgain-Pavlovi\'c \cite{BP} and Yoneda \cite{Yo} constructed some special periodic functions $u_0$
in some special critical Besov-Morrey space $X$ such that $\|u_0\|_{X}\leq \delta$.
They proved the norm of solution $u(t,x)$ satisfying that $\|u(T,x)\|_{X}\geq \delta^{-1}.$
Hence there is norm inflation phenomenon in the space $L^{\infty}([0,T),X)$.
It is easy to see that the quantity $\| e^{-\frac{t}{T\|u_0\|_{X}^2}} u(t,x)\|_{X}\leq \delta, \forall 0<t\leq T$.
That is to say, their results depend on the choice of solution spaces.

But in our paper, we proved that
$B_1(e^{t\Delta} u_1, e^{t\Delta } u_1) \notin {\mathcal{S}}'(\mathbb{R}^{n}).$
Our result can show blow-up phenomenon in the first step of the classic Picard iteration and
our ill-posedness result does not depend on the choice of solution spaces.
Hence our result is  a complement of the previous ill-posedness results on Navier-Stokes equations.

(ii) Cui \cite{Cui}  extended the results in \cite{BP} and \cite{Yo} to some logarithmic Besov spaces.
Cui's skills are based on refining the arguments of Wang \cite{W} and Yoneda \cite{Yo}.
All these ill-posedness results depend on the choice of solution spaces.

(iii) Besov-Morrey spaces can be found in \cite {LXY, LY, YSY}. In this paper, we consider only the critical Besov-Morrey spaces. Cui's Besov type spaces are logrithmically refined Besov space which are not critical spaces. Further, our skills can be applied also to Cui's spaces.

(iv) We don't impose any restriction on solution spaces in Theorem \ref{mthmain}.
Our initial value  satisfies $(1-\phi(x-\frac{3}{2}\vec{e}))u_0 \in ({\mathcal{S}}(\mathbb{R}^{n}))^{n}$, which  is located at the original and should have global weak solution.
But our result shows that we can not apply the classical Picard's process to get any solution for any $t>0$.
\end{remark}

To consider the blow-up phenomena of nonlinear term, we have made many preparations.
\begin{remark}
In \cite{LXY}, \cite{LY} and \cite{LYY}, we have considered semigroup structure and bilinear structure of non-linear terms.
In \cite{YY}, we have introduced parameter wavelet to control the influence of low frequency.
In \cite{YZ}, Yang-Zhu considered the multiplier operators and showed how the product of two functions produce the blow-up phenomena.
\end{remark}

This paper classify all the critical Besov-Morrey spaces into two class of spaces:
\begin{remark} By \cite{LXY}, $\dot{B}^{\gamma-1,\gamma}_{p,q}(1\leq p<\infty, 1\leq q\leq \infty, 0<\gamma\leq \frac{n}{p})$ have well-posedness property. By \cite{KT}, ${\rm BMO}^{-1}$ has well-posedness. Hence,
by Lemma \ref{le:2.7}, $\dot{B}^{-1,q}_{\infty}(q\leq 2)$ or $\dot{B}^{-1,0}_{2,q'}(q'\leq 2)$ or $\dot{B}^{-1,0}_{p,p'}(p<2,p'\geq 1)$ all have well-posedness property. That is to say,
{\bf for all the critical Besov-Morrey spaces } $\dot{B}^{\gamma-1,\gamma}_{p,q}(1\leq p<\infty, 1\leq q\leq \infty, 0\leq \gamma\leq \frac{n}{p})$,
according to Lemma \ref{le:2.7} in the next section, {\bf only the three classes of spaces in our Theorem \ref{mthmain} cause ill-posedness.}
\end{remark}

By the equation  (\ref{eq:two}), the value of left hand of the equation (\ref{eq:main}) is some multiple of the integration of correlation functions $h_{s,t}$ defined in (\ref{eq:hst}). We will use wavelets, vaguelets and the special property of Gauss function to prove the Theorem \ref{mthmain}.

The rest of this paper is organized as follows: In section 2, we will present some preliminaries about Meyer wavelets, vaguelets and Calder\'on-Zygmund operators, and we will further construct some special functions in Besov-Morrey spaces. In section 3, we will compute some integration related to Gauss functions and present some properties of two flows related to the equation (\ref{eq:main}). Finally, we will prove Theorem \ref{mthmain} in Section 4.

\section{Wavelets, special functions and operators}\label{sec2}

\subsection{Meyer wavelets} First of all,
we indicate that we will use tensorial product real valued
orthogonal Meyer wavelets. We refer the reader to \cite{Me, Woj, Yang1} for further information.
Let $\Psi^{0}$ be an even function in $ C^{\infty}_{0}
([-\frac{4\pi}{3}, \frac{4\pi}{3}])$ with
\begin{equation}
\left\{ \begin{aligned}
&0\leq\Psi^{0}(\xi)\leq 1; \nonumber\\
&\Psi^{0}(\xi)=1\text{ for }|\xi|\leq \frac{2\pi}{3}.\nonumber
\end{aligned} \right.
\end{equation}
  Write
  $$\Omega(\xi)= \sqrt{(\Psi^{0}(\frac{\xi}{2}))^{2}-(\Psi^{0}(\xi))^{2}}.$$ Then $\Omega(\xi)$ is an even
function in $ C^{\infty}_{0}([-\frac{8\pi}{3}, \frac{8\pi}{3}])$.
Clearly,

\begin{equation}
\left\{ \begin{aligned}
&\Omega(\xi)=0\text{ for }|\xi|\leq \frac{2\pi}{3};\nonumber\\
&\Omega^{2}(\xi)+\Omega^{2}(2\xi)=1=\Omega^{2}(\xi)+\Omega^{2}(2\pi-\xi)\text{
for }\xi\in [\frac{2\pi}{3},\frac{4\pi}{3}].\nonumber
\end{aligned} \right.
\end{equation}

 Let $\Psi^{1}(\xi)=
\Omega(\xi) e^{-\frac{i\xi}{2}}$. For any $\epsilon=
(\epsilon_{1},\cdots, \epsilon_{n}) \in \{0,1\}^{n}$, define
$\Phi^{\epsilon}(x)$ by $\hat{\Phi}^{\epsilon}(\xi)=
\prod\limits^{n}_{i=1} \Psi^{\epsilon_{i}}(\xi_{i})$. For $j\in
\mathbb{Z}$ and $k\in\mathbb{Z}^{n}$, let $\Phi^{\epsilon}_{j,k}(x)=
2^{\frac{nj}{2}} \Phi^{\epsilon} (2^{j}x-k)$. $\forall \epsilon\in \{0,1\}^{n}, j\in \mathbb{Z}, k\in \mathbb{Z}^{n}$ and distribution $f(x)$, denote
$f^{\epsilon}_{j,k}=\langle f, \Phi^{\epsilon}_{j,k}\rangle$.
Furthermore, we put
\begin{equation}\nonumber
\Lambda_{n} =\{(\epsilon,j,k), \epsilon\in \{0,1\}^{n}\backslash\{0\}, j\in\mathbb{Z},
k\in \mathbb{Z}^{n}\}.
\end{equation}
Then, the following result is well-known.
\begin{lemma}\label{le1}
The Meyer wavelets $\{\Phi^{\epsilon}_{j,k}(x)\}_{(\epsilon,j,k)\in
\Lambda_{n}}$ form an  orthogonal basis in $L^{2}(\mathbb{R}^{n})$.
Consequently, for any $f\in L^{2}(\mathbb{R}^{n})$, the following wavelet
decomposition holds in the $L^2$ convergence sense:
$$\begin{array}{c}
f(x)=\sum\limits_{(\epsilon,j,k)\in\Lambda_{n}}f^{\epsilon}_{j,k}\Phi^{\epsilon}_{j,k}(x).
\end{array}$$
\end{lemma}

\subsection{Properties of Besov-Morrey spaces}

We recall first wavelet characterization of oscillation spaces (see \cite{LXY,LY,YSY}).
Denote  $\mathfrak{D}=\{Q_{j,k}=2^{-j}k+2^{-j}[0,1]^{n}, \forall j\in \mathbb{Z}, k\in \mathbb{Z}^{n}\}$.
We have
\begin{lemma}\label{lem:c}
Given $1\leq p<\infty, 1\leq q \leq \infty$ and $\gamma_{1}\in \mathbb{R}, 0\leq \gamma_{2}\leq \frac{n}{p}$.

$(i)$\, $f(x)= \sum\limits_{\epsilon,j,k} a^{\epsilon}_{j,k}
\Phi^{\epsilon}_{j,k}(x)\in
\dot{B}^{\gamma_{1},0}_{\infty,q}(\mathbb{R}^{n})\Leftrightarrow$
\begin{equation}\label{eq:bc}
\begin{array}{rl}
&\Big[\sum\limits_{j}
2^{jq(\gamma_{1}+\frac{n}{2})}
\Big(\sup\limits_{(\epsilon,k)}
|a^{\epsilon}_{j,k}|\Big) ^{ q }\Big]^{\frac{1}{q}} <+
\infty,\end{array}
\end{equation}

$(ii)$\, $f(x)= \sum\limits_{\epsilon,j,k} a^{\epsilon}_{j,k}
\Phi^{\epsilon}_{j,k}(x)\in
\dot{B}^{\gamma_{1},\gamma_{2}}_{p,q}(\mathbb{R}^{n})\Leftrightarrow$
\begin{equation}\label{eq:bmc}
\begin{array}{rl}
&\sup\limits_{Q\in \mathfrak{D} }|Q|^{\frac{\gamma_{2}}{n}-\frac{1}{p}}
\Big\{\sum\limits_{nj\geq -\log_{2}|Q|}
2^{jq(\gamma_{1}+\frac{n}{2}-\frac{n}{p})}
\big(\sum\limits_{(\epsilon,k):Q_{j,k}\subset Q}
|a^{\epsilon}_{j,k}|^{p}\big) ^{ \frac{q}{p} }\Big\}^{\frac{1}{q}} <+
\infty.\end{array}
\end{equation}

\end{lemma}

Denote $\mathbb{N}=\{1,2,3,\cdots\}$. For $j\in \mathbb{N}$, take
$$h_{j}(x) = \sum\limits_{i=1,\cdots,n, 2^{j}\leq l_{i} \leq 2^{j+1}} \frac{2^{j}}{j^{\frac{1}{2}}} \Phi^{e}(2^{j} x-l)$$
and denote
\begin{equation}\label{eq:h}
h(x)=\sum\limits_{j\in \mathbb{N}} h_{j}(x).\end{equation}
By wavelet characterization Lemma \ref{lem:c}, we have
\begin{corollary}\label{cor.2.3}
$$h(x)\in \dot{B}^{-1,q}_{\infty} \bigcap \dot{B}^{-1,0}_{2,q'} \bigcap \dot{B}^{-1,0}_{p,p'} \, (\forall p,q,q'>2, p'\geq 1).$$
\end{corollary}

As a generalization of Morrey spaces, Besov-Morrey spaces cover many important function spaces, for example, Sobolev spaces, Besov spaces, Morrey
spaces, ${\rm BMO}^{-1}$, $Q$-spaces and so on. For an overview, we refer to Li-Xiao-Yang \cite{LXY}, 
Lin-Yang \cite{LY},Yang \cite{Yang1} and Yuan-Sickel-Yang \cite{YSY}.

\begin{lemma} \label{le:2.3}
$(i)$\, If $1\leq p<\infty, 1\leq q\leq \infty, \gamma_1\in \mathbb{R}$, $\gamma_2=\frac{n}{p}$, then the above Besov-Morrey spaces $\dot{B}^{\gamma_{1},\gamma_{2}}_{p,q}$  become the relative Besov spaces
$\dot{B}^{\gamma_{1},q}_{p}$.

$(ii)$\, If $p=\infty$ and $1\leq q\leq \infty$, then $\dot{B}^{\gamma_{1},\gamma_{2}}_{\infty,q}(\mathbb{R}^{n})=\dot{B}^{\gamma_{1},q}_{\infty}(\mathbb{R}^{n})$.

$(iii)$\, For $0<\alpha< \min ( \frac{n}{2}, 1 )$, $\dot{B}^{\alpha,\alpha}_{2,2}=Q_{\alpha}$.

\end{lemma}

Critical spaces occupied a significant place for Navier-Stokes equations (\ref{eqn:ns}).
If $u(t,x)$ is a solution of (\ref{eqn:ns}) with initial value $u_0(x)$, we replace $u(t,x)$, $p(t,x)$ and $u_0(x)$ by $u_{\lambda}(t,x)= \lambda u(\lambda^2 t, \lambda x), p_{\lambda}(t,x)= \lambda^2 p(\lambda^2 t,tx)$ and $u^0_{\lambda} (x) =\lambda u_0(\lambda x)$, respectively. Then $u_{\lambda}(t,x)$ is a solution of (\ref{eqn:ns}) with initial value $u_{\lambda}^{0}(x)$.
\begin{definition}
If $\|u_0(x)\|_{X}\sim \|u^{0}_{\lambda}(x)\|_{X}$ for any $\lambda>0$, then $X$ is called to be a critical space.
\end{definition}
For the function spaces defined above, if $\gamma_1-\gamma_2=-1$, then they are critical spaces.
\begin{lemma} \label{le:2.5}
For $1\leq p,q\leq \infty$ and $0\leq \gamma_2\leq \frac{n}{p}$,
$\dot{B}^{\gamma_2-1,\gamma_2}_{p,q}$  are critical spaces.
\end{lemma}

For $1\leq p,q\leq \infty$, the critical Besov-Morrey spaces have the following inclusion relation:
\begin{lemma}\label{lem:2.9}
Given $1\leq p, q\leq \infty, 0\leq \gamma'_2\leq \gamma_2\leq \frac{n}{p}$.
$$\dot{B}^{\frac{n}{p}-1, q}_{p}\subset \dot{B}^{\gamma_{2}-1, \gamma_{2}}_{p,q} \subset \dot{B}^{\gamma'_{2}-1, \gamma'_{2}}_{p,q}\subset \dot{B}^{-1,\infty}_{\infty}.$$
\end{lemma}

The above lemma shows that $\dot{B}^{-1,0}_{p,q}  (1\leq p,q\leq \infty)$ are the relative large spaces among the critical Besov-Morrey spaces.
For critical Besov-Morrey spaces $\dot{B}^{-1,0}_{p,q}(1\leq p,q\leq \infty)$, by their wavelet characterization in Lemma \ref{lem:c}, we have
\begin{lemma} \label{le:2.7}
$$\bigcup\limits_{q\leq 2} \dot{B}^{-1,q}_{\infty}\bigcup\limits_{q'\leq 2} \dot{B}^{-1,0}_{2,q'}\bigcup\limits_{p<2,p'\geq 1}\dot{B}^{-1,0}_{p,p'}\subset
{\rm BMO}^{-1}\subset \bigcap\limits_{q>2} \dot{B}^{-1,q}_{\infty}\bigcap\limits_{q'>2} \dot{B}^{-1,0}_{2,q'}\bigcap\limits_{p>2,p'\geq 1}\dot{B}^{-1,0}_{p,p'} .$$
\end{lemma}

\subsection{Vaguelets and Calder\'on-Zygmund operators}

Now we introduce some preliminaries on Calder\'on-Zygmund
operators, see \cite{Me} and \cite{Stein2}. For $x\neq y$, let
$K(x,y)$ be a smooth function such that
\begin{equation}\label{eq2}
|\partial ^{\alpha}_{x}\partial ^{\beta}_{y} K(x,y)| \leq
\frac{C}{|x-y|^{n+|\alpha|+|\beta|}}, \forall |\alpha|+ |\beta|\leq
N_{0}.
\end{equation}
In this paper, we assume that  $N_{0}$ is a large enough constant.
\begin{definition}
A linear operator $T$ is said to be a Calder\'on-Zygmund operator in $CZO(N_{0})$
if
\begin{itemize}
\item[(1)]
$T$ is continuous from $C^{1}(\mathbb{R}^{n})$ to $(C^{1}(\mathbb{R}^{n}))'$;
\item[(2)] There exists a kernel $K(\cdot,\cdot)$ satisfying (\ref{eq2}) and
for $x\notin {\rm supp} f$,
$$Tf(x)=\int K(x,y) f(y) dy;$$
\item[(3)] $Tx^{\alpha}=T^{*}x^{\alpha}=0, \forall \alpha \in \mathbb{N}^{n}$
and $|\alpha|\leq N_{0}-1$.\end{itemize}
\end{definition}

According to Schwartz kernel theorem, the kernel $K(x,y)$
of a linear continuous operator $T$ is only a distribution in $S'(\mathbb{R}^{2n})$.
Meyer-Yang \cite{MY} proved the continuity of Calder\'on-Zygmund operators on Besov spaces and Triebel-Lizorkin spaces. Lin-Yang \cite{LY} considered the relative continuity on Besov-Morrey spaces. In fact, we have
\begin{lemma}\label{CZ:con}
Given $1\leq p,q \leq \infty$ and $T\in CZO(2)$, then
\begin{equation*} T \mbox { is continuous from } \dot{B}^{-1,0}_{p,q} \mbox{ to }\dot{B}^{-1,0}_{p,q}. \end{equation*}
\end{lemma}

In this paper, we will use Calder\'on-Zygmund operators generated by vagulettes.
Denote $\psi^{1}(x)=x_2 x_3 e^{-x^{2}}$ and $\psi^{2}(x) = -x_1 x_3 e^{-x^{2}}$. We have $$\int x^{\alpha}\psi^{i}(x) dx=0, \forall i=1,2, 0\leq |\alpha|\leq 1.$$
Denote $\psi^{i}_{j,k}(x)= 2^{\frac{nj}{2}}\psi^{i}(2^{j}x-k)$. Then $\{\psi^{i}_{j,k}\}_{j\in \mathbb{Z},k\in \mathbb{Z}^{n}}$ are vaguelets generated by the derivatives of Gauss functions. The definition of vaguelets can be found in Meyer \cite{Me}.
For $i=1,2$, define
\begin{equation}\label{eq:T}
\left\{\begin{array}{ll}
T_{i}: \Phi^{\epsilon}_{j,k}(x) = \psi^{i}_{j,k}(x),& \mbox{ if } \epsilon=e;\\
T_{i}: \Phi^{\epsilon}_{j,k}(x) =0,& \mbox{ if } \epsilon \neq e.
\end{array}
\right.
\end{equation}
It is easy to see
\begin{lemma}\label{le:2.11}
For $n\geq 3$, we have

$(i)$\, $T_{i}\in CZO(2), \forall i=1,2$;

$(ii)$\, the divergence of $(T_1 h, T_2h,0, \cdots, 0)^{t}$ is zero.
\end{lemma}

By Corollary \ref{cor.2.3} and Lemma \ref{CZ:con}, we have
\begin{corollary} \label{le:2.12}
$$T_{i}h(x)\in \dot{B}^{-1,q}_{\infty} \bigcap \dot{B}^{-1,0}_{2,q'} \bigcap \dot{B}^{-1,0}_{p,p'}\,\quad \forall\, i=1,\,2,\, p,\,q,\,q'>2,\, p'\geq 1.$$
\end{corollary}

For $i=1,2$, $T_{i}h(x)$ are distributions concentrated on the cube $[1,2]^{n}$ and $T_{i}f(x)$ are good functions outside of the cube $[\frac{1}{2}, \frac{5}{2}]^n$. Particulary,
\begin{lemma}\label{le:2.13}
\begin{equation}\label{eq:THS}
(1-\phi(x-\frac{3}{2}\vec{e})) T_{i}h(x)\in {\mathcal{S}} (\mathbb{R}^{n}).
\end{equation}
\end{lemma}

\begin{proof}

Denote $x^{2}=\sum\limits^{n}_{i=1}x_{i}^2$ and for $\gamma\in \mathbb{N}^{n}$, denote $x^{\gamma}= \prod\limits^{n}_{i=1} x_{i}^{\gamma_i}$. For $j\geq 1$ and $i=1,\cdots,n$, $2^{j}\leq l_{i}\leq 2^{j+1}$, we have
\begin{equation}\label{eq:le}
(2^{-j}l-\frac{3}{2}\vec{e})^{2}\leq \frac{n}{4}.
\end{equation}
If $(x-\frac{3}{2}\vec{ e})^{2}\geq 4n$, then
\begin{equation}\label{eq:xl}
(x-2^{-j}l)^{2}\geq \frac{1}{2} (x-\frac{3}{2}\vec{ e})^{2} - (2^{-j}l -\frac{3}{2} \vec{e})^{2}\geq 2n -\frac{n}{4}\geq \frac{7n}{4};
\end{equation}
\begin{equation}\label{eq:xe}
(x-\frac{3}{2}\vec{e})^{2} \leq 2(x-2^{-j}l)^{2}+ 2(2^{-j}l-\frac{3}{2}\vec{e})^{2} \leq 3(x-2^{-j}l)^{2}.
\end{equation}

For $\beta, \gamma\in \mathbb{N}^{n}$, we have
\begin{equation}\label{eq:bg1}
|\partial^{\beta}_{x} \{ (2^{j}x-l)^{\gamma} e^{-(2^{j}x-l)^{2}}\}|\leq C 2^{j|\beta|} (2^{j}x-l)^{|\beta|+|\gamma|} e^{-(2^{j}x-l)^{2}}.
\end{equation}

Combine the equations from (\ref{eq:le}) to (\ref{eq:bg1}), for any $N\geq 0$, $\alpha,\, \beta,\, \gamma\in \mathbb{N}^{n}$, we have
\begin{equation}\label{eq:Nabg}
\begin{array}{rl}
&|(x-\frac{3}{2}\vec{e})^{2N} \partial^{\alpha}_{x} (1-\phi(x-\frac{3}{2}\vec{e}))  \partial^{\beta}_{x} \{ (2^{j}x-l)^{\gamma} e^{-(2^{j}x-l)^{2}}\}|\\
&\qquad\leq C 2^{j(|\beta|-2N) } (2^{j}x-l)^{2N+|\beta|+|\gamma|} e^{-(2^{j}x-l)^{2}}\\
&\qquad\leq C 2^{-j(2N+n+2) } (2^{j}x-l)^{2N+2|\beta|+|\gamma|+n+2} e^{-(2^{j}x-l)^{2}}\\
&\qquad\leq C 2^{-j(2N+n+2) } .
\end{array}
\end{equation}

Hence, for any $N\geq 0$, $\alpha,\, \beta \in \mathbb{N}^{n}$, we have
\begin{equation}\label{eq:Nab}
\begin{array}{rl}
&|(x-\frac{3}{2}\vec{e})^{2N} \partial^{\alpha}_{x} (1-\phi(x-\frac{3}{2}\vec{e}))  \partial^{\beta}_{x} \{ T_i h(x)\}|\\
&\qquad\leq C \sum\limits_{j\in \mathbb{N}} 2^{-j(2N+n+2) } 2^{nj} 2^{j}
\leq C.
\end{array}
\end{equation}
The last equation implies the equation (\ref{eq:THS}).
\end{proof}

\section{Gauss function and two flows related}\label{sec4}

In this section, we first consider four integrations related to Gauss function, then we consider some properties of two flows which will be used to consider the blow up phenomenon in the next section.

\subsection{Four integrations relative to Gauss function}
In this subsection, we compute four integrations related to Gauss function and their derivatives which will be used to compute the expression of two flows in the next subsection.
For any $ 0<s< t$,  $j\in \mathbb{N}$, $x\in \mathbb{R}$ and $k\in \mathbb{Z}$, denote
$$I(x,t,s,0)= (t-s)^{-\frac{1}{2}}\int {\rm \, exp\, } \{- \frac{(x-y)^{2}}{4(t-s)} \} {\rm \, exp\, } \{-\frac{y^{2}}{4t}\} dy;$$
$$I(x,t,s,1)= (t-s)^{-\frac{3}{2}}\int (x_l-y_{l}) {\rm\, exp\, } \{- \frac{(x-y)^{2}}{4(t-s)} \} {\rm \, exp\, } \{-\frac{y^{2}}{4t}\} dy;$$
$$I(x,t,j,k,0)= t^{-\frac{1}{2}}\int {\rm \, exp\, } \{- \frac{(x-y)^{2}}{4t} \} {\rm \, exp\, } \{-|2^{j}y_l-k_{l}|^{2} \} dy;$$
$$I(x,t,j,k,1)= t^{-\frac{1}{2}}\int {\rm \, exp\, } \{- \frac{(x-y)^{2}}{4t} \} (2^{j}y-k)  {\rm \, exp\, } \{-|2^{j}y-k|^{2} \} dy.$$
Then we have
\begin{lemma}\label{le:3.1}
For $ 0\leq s\leq t, j\in \mathbb{N}, x\in \mathbb{R}, k\in\mathbb{Z}$, we have

$(i)$\, $I(x, t,s, 0)= 2\sqrt{\pi t}(2t-s)^{-\frac{1}{2}} {\rm \, \exp\, } \{-\frac{x^{2}}{4(2t-s)}\}$;

$(ii)$\, $I(x, t,s, 1)= 2\sqrt{\pi t}\, x (2t-s)^{-\frac{3}{2}} {\rm \, \exp\, } \{-\frac{x^{2}}{4(2t-s)}\}$;

$(iii)$\, $I(x, t,j,k, 0)= 2\sqrt{\pi } (1+4^{j+1}t)^{-\frac{1}{2}} {\rm \, \exp\, } \{-\frac{|2^{j}x-k|^{2}}{1+4^{j+1}t}\}$;

$(iv)$\, $I(x, t,j,k, 1)= 2\sqrt{\pi }(2^{j}x-k) (1+4^{j+1}t)^{-\frac{3}{2}} {\rm \, \exp\, } \{-\frac{|2^{j}x-k|^{2}}{1+4^{j+1}t}\}$.
\end{lemma}

\begin{proof}
(i) We regroup the function inside the integration and get
$$\begin{array}{rcl} I(x, t,s,0) &=& (t-s)^{-\frac{1}{2}} \int {\rm\, exp \,} \{-\frac{1}{4}(\frac{1}{t-s}+\frac{1}{t}) [ y-\frac{tx}{2t-s} ]^{2}\} dy \\
& & \times {\rm \, exp \,} \{ -\frac{x^{2}}{4(t-s)} + \frac{ tx^{2}}{ 4(t-s)(2t-s)}\}.\end{array}$$
Then, by changing variable, we have
$$\begin{array}{rcl} I(x, t,s,0) &=& (t-s)^{-\frac{1}{2}} \int {\rm\, exp \,} \{-\frac{1}{4}(\frac{1}{t-s}+\frac{1}{t})  y^{2}\} dy  {\rm \, exp \,} \{ -\frac{x^{2}}{4(2t-s)} \}\\
&=& (t-s)^{-\frac{1}{2}} \int {\rm\, exp \,} \{-\frac{2t-s}{4t(t-s)}  y^{2}\} dy  {\rm \, exp \,} \{ -\frac{x^{2}}{4(2t-s)} \}.\end{array}$$
Since $ \int {\rm\, \exp \,} \{-\frac{2t-s}{4t(t-s)}  y^{2}\} dy= 2\sqrt{\pi t} (t-s)^{\frac{1}{2}}  (2t-s)^{-\frac{1}{2}} $, we get
$$\begin{array}{rcl} I(x, t,s,0) &=& 2\sqrt{\pi t} (2t-s)^{-\frac{1}{2}}  {\rm \, exp \,} \{ -\frac{x^{2}}{4(2t-s)} \}.\end{array}$$

(ii) We make variable substitution
$z= y-\frac{tx}{2t-s}$. We have then $x-y= x-\frac{tx}{2t-s}-z$ and get
$$\begin{array}{rcl} I(x, t,s,1) &=& (t-s)^{-\frac{3}{2}} \int (\frac{(t-s)x}{2t-s}-z) {\rm\, exp \,} \{-\frac{1}{4}(\frac{1}{t-s}+\frac{1}{t}) z^{2}\} dz   {\rm \, exp \,} \{ -\frac{x^{2}}{4(2t-s)}  \}.\end{array}$$

Note that $ \int {\rm\, \exp \,} \{-\frac{1}{4}(\frac{1}{t-s}+\frac{1}{t})  z^{2}\} dz= 2\sqrt{\pi t} (t-s)^{\frac{1}{2}}  (2t-s)^{-\frac{1}{2}} $ and $ \int z {\rm\, exp \,} \{-\frac{1}{4}(\frac{1}{t-s}+\frac{1}{t})  z^{2}\} dz=0$, we obtain
$$\begin{array}{rcl} I(x, t,s,1) &=& 2\sqrt{\pi t} x (2t-s)^{-\frac{3}{2}}  {\rm \, exp \,} \{ -\frac{x^{2}}{4(2t-s)} \}.\end{array}$$

(iii) Regrouping the function inside the integration, we get
$$\begin{array}{rcl} I(x, t,j,k,0) &=& t^{-\frac{1}{2}} \int {\rm\, exp \,} \{-(\frac{1}{4t}+4^{j}) [ y-\frac{x+ 2^{j+2} t k}{1+4^{j+1}t} ]^{2}\} dy \\ & & \times {\rm \, exp \,} \{ (\frac{1}{4t}+4^{j}) \frac{(x+ 2^{j+2}t k)^{2}}{1+4^{j+1}t } - \frac{ x^{2}}{ 4t}- k^{2} \}.\end{array}$$
Then a variable substitution implies that
$$\begin{array}{rcl} I(x, t,j,k,0) &=& t^{-\frac{1}{2}} \int {\rm\, exp \,} \{-(\frac{1}{4t}+ 4^{j})  y^{2}\} dy  {\rm \, exp \,} \{ -\frac{|2^{j}x-k|^{2}}{1+4^{j+1}t} \}.\end{array}$$
Applying $ \int {\rm\, exp \,} \{-(\frac{1}{4t}+ 4^{j})  y^{2}\} dy= 2\sqrt{\pi t}  (1+ 4^{j+1}t )^{-\frac{1}{2}} $, we get
$$\begin{array}{rcl} I(x, t,j,k,0) &=& 2 \sqrt{\pi }(1+4^{j+1}t)^{-\frac{1}{2}}  {\rm \, exp \,} \{ -\frac{|2^{j}x-k|^{2}}{1+4^{j+1}t} \}.\end{array}$$

(iv) Regrouping the function inside the integration, we get
$$\begin{array}{rcl} I(x, t,j,k,1) &=& t^{-\frac{1}{2}} \int (2^{j}y-k) {\rm\, exp \,} \{-(\frac{1}{4t}+4^{j}) [ y-\frac{x+ 2^{j+2} t k}{1+4^{j+1}t} ]^{2}\} dy \\
& & \times {\rm \, exp \,} \{ (\frac{1}{4t}+4^{j}) \frac{(x+ 2^{j+2}t k)^{2}}{1+4^{j+1}t } - \frac{ x^{2}}{ 4t}- k^{2} \}.\end{array}$$
Further,
$2^{j}y-k = 2^{j}(y-\frac{x+ 2^{j+2} t k}{1+4^{j+1}t}) + (1+4^{j+1}t)^{-1} (2^{j}x-k) $.
Based on the fact $\int y {\rm\, \exp \,} \{ - (\frac{1}{4t}+ 4^{j}) y^{2} dy=0$, similar to the above (iii), we have
$$I(x, t,j,k, 1)= 2\sqrt{\pi }(2^{j}x-k) (1+4^{j+1}t)^{-\frac{3}{2}} {\rm \, exp\, } \{-\frac{|2^{j}x-k|^{2}}{1+4^{j+1}t}\}.$$
\end{proof}

For any $ t>0$,  $j\in \mathbb{N}$ and $x\in \mathbb{R}$, denote
$$I(x,t, j,0) = (4\pi)^{-\frac{1}{2}}\! \sum\limits_{k\in \mathbb{Z}} I(x,t, j,k,0) {\mbox {\rm \,  and  }}
I(x,t, j,1) = (4\pi)^{-\frac{1}{2}}\! \sum\limits_{k\in \mathbb{Z}} I(x,t, j,k,1).$$
We have
\begin {lemma} \label{le:3.4.1}

$(i)$\, Symmetry.\, $I(x,t,j,0)= I (3-x, t, j, 0).$

$(ii)$\, Anti-symmetry.\, $I(x,t,j,1)=- I(3-x, t,j,1).$

$(iii)$\, Zero point.\, $I(\frac{3}{2},t,j,1)=0.$

$(iv)$\, Positivity.\, $I(x,t,j,0)>0$ and if $x>\frac{3}{2}$, then $I(x,t,j,1)>0.$

$(v)$\, Negativity.\, $I(x,t,j,1)<0$, if $x<\frac{3}{2}.$
\end{lemma}

\begin{proof} Applying the expression in Lemma \ref{le:3.1}, we can get the above results. The details are omitted.
\end{proof}

\subsection{Expression of two flow functions}
In this subsection, we compute the expression of two flow functions.
The kernel of operator $e^{t\Delta}$ is $(4\pi t)^{-\frac{n}{2}} e^{-\frac{(x-y)^2}{4t}}$.
Let $A^{t}_{l}= \frac{\partial}{\partial x_l} e^{t\Delta}$.
We first consider the flow $A^{t-s}_{l} g(t,x)$.
The kernel of $A^{t-s}_{l}$ is
$$\frac{-(x_l-y_{l})}{2^{n+1} \pi^{\frac{n}{2}} (t-s)^{\frac{n}{2}+1}} {\rm \, exp\, } \{-\frac{(x-y)^{2}} {4t}\}.$$
By Lemma \ref{le:3.1}, we have
\begin{lemma} \label{le:3.2}
\begin{equation}\label{eq:AG}
A^{t-s}_{1} g(t,x) =   \frac{- t^{\frac{n}{2}} x_1 }{2(2t-s)^{\frac{n+2}{2}}}   {\rm \, exp \,} \{-\frac{x^{2}}{4(2t-s)}\}.
\end{equation}
\end{lemma}
\begin{proof}
$$ \begin{array}{rcl}
A^{t-s}_{1} g(t,x) &=& -\frac{2^{-n-1} \pi^{-\frac{n}{2}} }{\sqrt{t-s}} \int \frac{x_1-y_1}{t-s} {\rm \, exp \,} \{ -\frac{(x_1-y_1)^{2}}{4(t-s)} \} {\rm \, exp \,}
\{-\frac{y_1^2}{4t}\} dy_1\\
& & \times  \prod\limits^{n}_{l=2}\frac{1}{\sqrt{t-s}} \int  {\rm \, exp \,} \{ -\frac{(x_l-y_l)^{2}}{4(t-s)} \} {\rm \, exp \,}
\{-\frac{y_l^2}{4t}\} dy_l\\
 &=& -2^{-n-1} \pi^{-\frac{n}{2}} x_1 \frac{2 \sqrt{\pi t} }{(2t-s)^{\frac{3}{2}}}  {\rm \, exp \,} \{ -\frac{x_1^{2}}{4(2t-s)} \} \times  \frac{2^{n-1}(\pi t)^{\frac{n-1}{2}}}{(2t-s)^{\frac{n-1}{2}}}   \prod\limits^{n}_{l=2}  {\rm \, exp \,} \{-\frac{x_l^{2}}{4(2t-s)}\} \\
 &=&  \frac{-t^{\frac{n}{2}} x_1 }{2(2t-s)^{\frac{n+2}{2}}}   {\rm \, exp \,} \{-\frac{x^{2}}{4(2t-s)}\}.
\end{array}$$
\end{proof}

Next we consider another flow. Let \begin{equation}\label{eq:u1} v_{1}(x)= \sum\limits_{j\in \mathbb{N}} v_{1,j}(x), \mbox{ where } \end{equation}
$$v_{1,j}(x) = \sum\limits_{i=1,\cdots,n, 2^{j}\leq l_{i} \leq 2^{j+1}}  2^{j} j^{-\frac{1}{2}} \psi^{1}(2^{j}x-l).$$
The kernel of $e^{t\Delta}$ is $
(4\pi t)^{-\frac{n}{2}} {\rm \, \exp \,} \{-\frac{(x-y)^{2}}{4t}\}.$
By Lemma \ref{le:3.1}, we have
\begin{lemma} \label{le:3.3}
For $j \in \mathbb{N}$, we have
\begin{equation}\label{eq:vj}\begin{array}{rcl}e^{t\Delta} v_{1,j}(x)\!\! &\!=\!&\!\! (4\pi)^{-\frac{n}{2}}  2^{j} j^{-\frac{1}{2}}
I (x_1,t,j,0)\prod\limits^{3}_{l=2}I(x_l,t,j,1)\prod\limits^{n}_{l'=4} I (x_{l'},t,j,0) \\
\!\! &\!=\!&\!\!  \sum\limits_{i=1,\cdots,n, 2^{j}\leq k_{i} \leq 2^{j+1}}  2^{j} j^{-\frac{1}{2}}  \frac{(2^{j}x_2-k_2) (2^{j} x_3-k_3)} {(1+4^{j+1}t)^{\frac{n}{2}+2}}        {\rm \, exp \, }\{ - \frac {(2^{j}x-k)^2}{1+4^{j+1}t}\}.
\end{array}
\end{equation}
Hence,
\begin{equation}\label{eq:u1d}
\begin{array}{c}
e^{t\Delta} v_{1}(x) =\\
\sum\limits_{j\in \mathbb{N}}
\sum\limits_{i=1,\cdots,n, 2^{j}\leq k_{i} \leq 2^{j+1}}  2^{j} j^{-\frac{1}{2}}  \frac{(2^{j}x_2-k_2) (2^{j} x_3-k_3)} {(1+4^{j+1}t)^{\frac{n}{2}+2}}        {\rm \, exp \, }\{ - \frac {(2^{j}x-k)^2}{1+4^{j+1}t}\}.
\end{array}
\end{equation}
\end{lemma}

\begin{proof} The conclusions of Lemma \ref{le:3.3} are obtained by the following equality:
$$\begin{array}{c}
e^{t\Delta} \{ (2^{j}x_2-k_2) (2^{j} x_3-k_3)  {\rm \, exp \, }\{ - (2^{j}x-k)^2 \}\} \\
= \frac{(2^{j}x_2-k_2) (2^{j} x_3-k_3)} {(1+4^{j+1}t)^{\frac{n}{2}+2}}        {\rm \, exp \, }\{ - \frac {(2^{j}x-k)^2}{1+4^{j+1}t}\}.
\end{array}$$
\end{proof}

\subsection{ Some properties of flows}
In this subsection, we consider the following four properties of the above two flow functions:
symmetry, monotonicity, positivity and zero point.
To present well the symmetry properties of flow functions $A^{t-s}_{1} g(t,x)$,
we introduce the following notations:
$$x^{-,0}_{i}= (x_1, \cdots, x_{i-1}, -x_i, x_{i+1},\cdots, x_n).$$
For $A^{t-s}_{1} g(t,x)$, we have

\begin {lemma}

$(i)$\, Positivity.\, $A^{t-s}_{1} g(t,x)>0, \forall\, x_1>0;$

$(ii)$\, Anti-symmetry.\, $A^{t-s}_{1} g(t,x)= -A^{t-s}_{1} g(t, x^{-,0}_1);$

$(iii)$\, Symmetry.\, $A^{t-s}_{1} g(t,x)= A^{t-s}_{1} g(t, x^{-,0}_i), \forall i=2,\cdots, n.$
\end{lemma}

To present well the properties of flow functions $e^{t\Delta} u_{1,j}(x)$,
we introduce the following notations:
$$x^{-,3}_{i}= (x_1, \cdots, x_{i-1}, 3-x_i, x_{i+1},\cdots, x_n),$$
$$x^{0}_{i}= (x_1, \cdots, x_{i-1}, \frac{3}{2}, x_{i+1},\cdots, x_n).$$
For $e^{t\Delta} v_{1,j}(x)$, we have

\begin {lemma} \label{le:3.4}

$(i)$\, Anti-symmetry.\, $e^{t\Delta} v_{1,j}(x) = - e^{t\Delta} v_{1,j}(x^{-,3}_{i}), \forall i=2,3.$

$(ii)$\, Symmetry.\, $e^{t\Delta} v_{1,j}(x)= e^{t\Delta} v_{1,j}(x^{-,3}_{i}), \forall i=1,4,\cdots,n.$

$(iii)$\, Zero point.\, $e^{t\Delta} v_{1,j}(x^{0}_{2})=e^{t\Delta} v_{1,j}(x^{0}_{3})=0.$

$(iv)$\, Positivity.\, $e^{t\Delta} v_{1,j}(x)>0$, if $(x_2-\frac{3}{2})(x_3-\frac{3}{2})>0.$

$(v)$\, Negativity.\, $e^{t\Delta} v_{1,j}(x)<0$, if $(x_2-\frac{3}{2})(x_3-\frac{3}{2})<0.$
\end{lemma}

\begin{proof} Applying the properties in Lemma \ref{le:3.4.1}
and applying the expression of flow functions in Lemma \ref{le:3.2},
we can get the above results. The details are omitted.
\end{proof}

For $x=(x_1,x_2,\cdots, x_n)\in \mathbb{R}^{n}$,  denote
$x_{1}'= (x_2,\cdots, x_n)$  and  $\tilde{x} = (3+x_1, x_{1}')$. 
For $t>0$, $v_1(x)$ defined in (\ref{eq:u1}) and $ e^{t\Delta} v_1(x)$ defined in (\ref{eq:u1d}), denote
$V(t,x) = (e^{t\Delta} v_{1}(x))^{2}$. 
For $j\in \mathbb{N}$,  denote
$$I (x_{1}',t,j)=  2^{j} j^{-\frac{1}{2}} \prod\limits^{3}_{l=2}I(x_l,t,j,1)\prod\limits^{n}_{l'=4} I (x_{l'},t,j,0).$$
Then

\begin{lemma}  \label{le:3.34}
For $t>0, j\in \mathbb{N}, x_1 \geq 0,\, x_{1}'\in \mathbb{R}^{n-1}$, we have
\begin{equation}\label{eq:vm0} I (x_{1},t,j,0)\geq I (x_{1}+3,t,j,0).\end{equation}
\begin{equation}\label{eq:vm}|e^{t\Delta} v_{1,j}(x)|\geq |e^{t\Delta} v_{1,j}(\tilde{x})|.\end{equation}
Hence,
\begin{equation}\label{eq:Vm}V(t,x)-V(t,\tilde{x})\geq 0.\end{equation}

\end{lemma}

\begin{proof}
(i) For $j\in \mathbb{N}, x_1\geq 0$ and $2^{j}\leq k_1\leq 2^{j+1}$, we have
$$3\cdot 2^{2j}+ 2(2^{j} x_1-k_1)\geq 3\cdot 2^{2j}-2k_1\geq 3\cdot 2^{2j}- 2^{j+2}\geq 0.$$
That is to say,
$$(2^{j}(x_1+3)-k_1)^{2}\geq (2^{j}x_1-k_1)^{2}.$$
Hence we get \eqref{eq:vm0}.

(ii) By applying Lemmas \ref{le:3.4.1} and \ref{le:3.4}, according to \eqref{eq:vm0}, we have
$$|e^{t\Delta} v_{1,j}(x)|\geq |e^{t\Delta} v_{1,j}(\tilde{x})|.$$

(iii) Observe that
\begin{equation} \label{eq:VV}
\begin{array}{rl} & V(t,x)- V(t,\tilde{x})\\
=& \sum\limits_{j,j'\in \mathbb{N}} I(x_1, t, j,0)I(x_1, t, j',0)I(x_{1}', t, j)I(x_{1}', t, j') \\
- & \sum\limits_{j,j'\in \mathbb{N}} I(3+x_1, t, j,0)I(3+x_1, t, j',0)I(x_{1}', t, j)I(x_{1}', t, j')\\
=& \sum\limits_{j,j'\in \mathbb{N}} \{I(x_1, t, j,0)I(x_1, t, j',0) - I(3+x_1, t, j,0)I(3+x_1, t, j',0)\} \\
&\times I(x_{1}', t, j)I(x_{1}', t, j').
\end{array}
\end{equation}
By \eqref{eq:vm0}, we have
\begin{equation} \label{eq:VV1}
I(x_1, t, j,0)I(x_1, t, j',0) - I(3+x_1, t, j,0)I(3+x_1, t, j',0)\geq 0.
\end{equation}
By Lemma \ref{le:3.4.1}, we have
\begin{equation} \label{eq:VV2}
I(x_{1}', t, j)I(x_{1}', t, j')\geq 0.
\end {equation}
By equations \eqref{eq:VV}, \eqref{eq:VV1} and \eqref{eq:VV2}, we get \eqref{eq:Vm} is true.

\end{proof}

\subsection{ Correlation function and non-integrability}
Let $h_{s,t}$ be the correlation function between function $V(s,x)$ and function $A^{t-s}_{1}g(t,x)$, i.e.,
\begin{equation}\label{eq:hst}h_{s,t} = -2 \cdot t^{-\frac{n}{2}} \langle V(s,x), A^{t-s}_{1} g(t,x)\rangle .\end{equation} Then we have
\begin{theorem} \label{th3.05}
$\forall\, 0<s<t<1, h_{s,t} \geq 0.$
\end{theorem}
\begin{proof}
Denote $x_{1}'= (x_2,\cdots, x_n)$ and $x^{-,0}_{i}= (-x_1, x_{2},\cdots, x_n).$ Then
$$\begin{array}{rl}
h_{s,t}= &   \int^{\infty}_{0} \int_{\mathbb{R}^{n-1}} V(s,x) \frac{x_1}{(2t-s)^{\frac{n}{2}+1}} {\rm \, exp \, } \{ -\frac{x^{2}}{4 (2t-s)}\} dx_1 dx'_{1}\\
& - \int^{\infty}_{0} \int_{\mathbb{R}^{n-1}} V(s, x^{-,0}_{1}) \frac{x_1}{(2t-s)^{\frac{n}{2}+1}} {\rm \, exp \, } \{ -\frac{x^{2}}{4 (2t-s)}\} dx_1 dx'_{1}\\
= &  \int^{\infty}_{0} \int_{\mathbb{R}^{n-1}} [V(s,x)-V(s, \tilde{x})] \frac{x_1}{(2t-s)^{\frac{n}{2}+1}} {\rm \, exp \, } \{ -\frac{x^{2}}{4 (2t-s)}\} dx_1 dx'_{1}\\
= &  \int^{\frac{3}{2}}_{0} \int_{\mathbb{R}^{n-1}} [V(s,x)-V(s,\tilde{x})] \frac{x_1}{(2t-s)^{\frac{n}{2}+1}} {\rm \, exp \, } \{ -\frac{x^{2}}{4 (2t-s)}\} dx_1 dx'_{1}\\
&+  \int^{\infty}_{\frac{3}{2} } \int_{\mathbb{R}^{n-1}} [V(s,x)-V(s,\tilde{x})] \frac{x_1}{(2t-s)^{\frac{n}{2}+1}} {\rm \, exp \, } \{ -\frac{x^{2}}{4 (2t-s)}\} dx_1 dx'_{1}
\end{array}$$
By the equation (\ref{eq:Vm}) of $V(s,x)$, we have
\begin{equation}\label{eq:h36}
\begin{array}{rl}
& h_{s,t} \\
\geq \!\!\!  &  \! \int^{3}_{\frac{3}{2} } \int_{\mathbb{R}^{n-1}} [V(s,x)-V(s, \tilde{x})] \frac{x_1}{(2t-s)^{\frac{n}{2}+1}} {\rm \, exp \, } \{ -\frac{x^{2}}{4 (2t-s)}\} dx_1 dx'_{1}\\
\geq  \!\!\!  &  \!\int_{[2,3]^{n}} [V(s,x)-V(s, \tilde{x})] \frac{x_1}{(2t-s)^{\frac{n}{2}+1}} {\rm \, exp \, } \{ -\frac{x^{2}}{4 (2t-s)}\} dx_1 dx'_{1}\\
\geq \!\!\! & 0.
\end{array}
\end{equation}
\end{proof}

\begin{theorem} \label{th3.15} For any $j\in \mathbb{N}, 0<s<t<1$ and $2^{-2j}\leq s<2^{2-2j}$, there exists $c>0$ such that
$$h_{s,t} \geq  \frac{c 2^{2j}}{j}.$$
\end{theorem}
\begin{proof}

By (\ref{eq:h36}),
$$\begin{array}{rl}
h_{s,t}\geq   & \int_{[2,3]^{n}} [V(s,x)-V(s, x_1 +3, x_{1}')] \frac{x_1}{(2t-s)^{\frac{n}{2}+1}} {\rm \, exp \, } \{ -\frac{x^{2}}{4 (2t-s)}\} dx_1 dx'_{1}.
\end{array}$$
By equations (\ref{eq:vm}), \eqref{eq:VV}, \eqref{eq:VV1} and \eqref{eq:VV2}, for $j\in \mathbb{N}$ and $x\in [2,3]^{n}$, we have
$$V(s,x)-V(s, \tilde{x})\geq (e^{s\Delta}v_{1,j}(x))^{2}- (e^{s\Delta}v_{1,j}(\tilde{x}))^{2}.$$
Further, if $2^{-2j}\leq s< 2^{2-2j}$ and $x\in [2,3]^{n}$, then
$$(e^{s\Delta}v_{1,j}(x))^{2}- (e^{s\Delta}v_{1,j}(\tilde{x}))^{2}\geq \frac{1}{2} (e^{s\Delta}v_{1,j}(x))^{2}\geq \frac{ c 2^{2j}}{j}.$$
Hence, for $2^{-2j}\leq s< 2^{2-2j}$ and $x\in [2,3]^{n}$,
$$[V(s,x)-V(s, \tilde{x})]_{[2,3]^{n}}\geq \frac{c 2^{2j}}{j}.$$
Consequently,
$$ h_{s,t}\geq   \frac{c 2^{2j}}{j}.$$
\end{proof}

\begin{theorem} \label{th3.5}
$\int^{t}_{0} h_{s,t} ds=\infty, \forall\, 0<t<1.$
\end{theorem}
\begin{proof}
For any $0<t<1$, there exists $j_{t}\in \mathbb{N}$ such that $2^{-2j_{t}}\leq t< 2^{2-2j_{t}}$.
Then,
$$\int^{t}_{0} h_{s,t} ds\geq \sum\limits_{j\geq j_{t}} \frac{c 2^{2j}}{j} 2^{-2j} =  \sum\limits_{j\geq j_{t}} \frac{c }{j}=+\infty.$$
\end{proof}

\section{Proof of Theorem 1.2}
\begin{proof} We define $u_0$ as follows:
\begin{equation} \label{eq:u0} u_{0} = (u_{1}, u_{2}, u_{3}, \cdots, u_{n})^{t}= (\delta T_{1}h, \delta T_{2}h, 0, \cdots,0)^{t}.
\end{equation}
According to Lemmas \ref{le:2.11}, \ref{le:2.13} and Corollary \ref{le:2.12}, we have
\begin{itemize}
\item[(i)] ${\rm \, div\,} u_{0}=0$;
\item[(ii)] for any $p,\,q,\,q'>2,\, p'\geq 1$,
$$\max\{\|u_0\|_{(\dot{B}^{-1,q}_{\infty})^{n}}, \|u_0\|_{(\dot{B}^{-1,0}_{2,q'})^{n}}, \|u_0\|_{(\dot{B}^{-1,0}_{p,p'})^{n}}, \|u_0\|_{(\dot{F}^{-1,0}_{p',p})^{n}}\}\leq c\delta;$$
\item[(iii)] $(1-\phi(x-\frac{3}{2}\vec{e})) u_{0}\in \big({\mathcal{S}} (\mathbb{R}^{n})\big)^{n}.$
\end{itemize}
By definition of $B_1$, we get
\begin{equation}\label{eq:two}
\langle B_{1}(e^{t\Delta}u_{1}, e^{t\Delta} u_{1}), g(t,x)\rangle = - \int ^{t}_{0} \langle (e^{s\Delta} u_1)^2, \frac{\partial}{\partial x_{1}} e^{(t-s)\Delta}g(t,x)\rangle ds.
\end{equation}
Hence,
$$\langle B_{1}(e^{t\Delta}u_{1}, e^{t\Delta} u_{1}), g(t,x)\rangle  = 2^{-1} \delta^{2}  t^{\frac{n}{2}} \int ^{t}_{0} h_{s,t} ds.$$
Invoking Theorem \ref{th3.5} leads to
$$\langle B_{1}(e^{t\Delta}u_{1}, e^{t\Delta} u_{1}), g(t,x)\rangle  =  \infty.$$
This completes the proof of Theorem 1.2.
\end{proof}

{\bf Acknowledgement:}
The authors would like thank professor Yves Meyer and professor Chaojiang Xu for their interesting of this work and many useful advices.


\end{document}